\documentclass[twocolumn,showpacs,preprintnumbers,amsmath,amssymb]{revtex4}

\usepackage{graphicx}

 \newcommand\la{\langle}
 \newcommand\ra{\rangle}
 \newcommand\beq{\begin{equation}}
 
 \newcommand\eeq{\end{equation}}
 \newcommand\beqn{\begin{eqnarray}}
 \newcommand\eeqn{\end{eqnarray}}

\def\sq{\sigma^N_{\bar qq}}
\def\mb{\,\mbox{mb}}
\def\fm{\,\mbox{fm}}
\def\GeV{\,\mbox{GeV}}

\def\lsim{\mathrel{\rlap{\lower4pt\hbox{\hskip1pt$\sim$}}
    \raise1pt\hbox{$<$}}}         
\def\gsim{\mathrel{\rlap{\lower4pt\hbox{\hskip1pt$\sim$}}
    \raise1pt\hbox{$>$}}}         

\begin{document}

\title{Breakdown of QCD factorization at large Feynman \boldmath$x$}

\author{B.Z.~Kopeliovich$^{a,b}$}
\author{J.~Nemchik$^c$}
\author{I.K.~Potashnikova$^{a,b}$}
\author{M.B.~Johnson$^d$}
\author{Ivan~Schmidt$^a$}

\affiliation{$^a$Departamento de Fisica, Universidad Tecnica
Santa Maria, Valparaiso, Chile\\
$^b$Max-Planck Institut f\"ur Kernphysik, Heidelberg, Germany\\
$^c$Institute of Experimental Physics SAV, Kosice, Slovakia\\
$^d$Los Alamos National Laboratory, Los Alamos, NM 87545, USA}

\date{\today}

\begin{abstract}

 Recent measurements by the BRAHMS collaboration of high-$p_T$ hadron
production at forward rapidities at RHIC found the relative production
rate $(d-Au)/(p-p)$ to be suppressed, rather than enhanced. Examining
other known reactions (forward production of light hadrons, the Drell-Yan
process, heavy flavor production, etc.), one notes that all of these
display a similar property, namely, their cross sections in nuclei are
suppressed at large $x_F$.  Since this is the region where $x_2$ is
minimal, it is tempting to interpret this as a manifestation of coherence,
or of a color glass condensate, whereas it is actually a simple
consequence of energy conservation and takes place even at low energies.
We demonstrate that in all these reactions there is a common suppression
mechanism that can be viewed, alternatively, as a consequence of a reduced
survival probability for large rapidity gap processes in nuclei, Sudakov
suppression, an enhanced resolution of higher Fock states by nuclei, or an
effective energy loss that rises linearly with energy. Our calculations
agree with data.

\end{abstract}


\maketitle

\section{Introduction}

It is believed that in order to reach the smallest values of the
light-front momentum fraction variable $x$ in nuclei, and thus access the
strongest coherence effects such as those associated with shadowing or the
color glass condensate, one should go to forward rapidities, i.e., to the
beam fragmentation region at large Feynman $x_F$. This probably was the
idea behind the measurements of high-$p_T$ hadron production in
deuteron-gold collisions at large rapidities performed recently by the
BRAHMS collaboration \cite{brahms} at RHIC.

This proposition is based upon the usual leading order relation between
$x_1$ and $x_2$ for two colliding partons, $x_1\,x_2=\hat s/s$, where
$\hat s$ and $s$ are the square of the c.m. energies of the colliding
partons and hadrons, respectively. It is demonstrated in
Sect.~\ref{shadowing}, however, that although formally $x_2$ reaches its
minimal values as $x_1\to 1$, the coherence phenomena vanish in this limit
\cite{e-loss}.

Moreover, it is shown in Sect.~\ref{sudakov} that another effect causing
considerable nuclear suppression for any reaction at large $x_F$ can be
easily misinterpreted as coherence. The source of this suppression may be
understood in terms of the Sudakov form factor, which is the probability
that no particles be produced as $x_F\to 1$, as demanded by energy
conservation. Clearly, multiple interactions make it more difficult to
satisfy this condition and therefore should cause an even greater
suppression.

Spectator partons, both soft and hard, may also interact while propagating
through the nucleus and populate the large rapidity gap (LRG) that forms
when $x_F\to 1$. As long as production within the LRG is forbidden, the
presence of spectators further reduces the survival probability of LRG
processes \cite{maor}.

This mechanism of suppression can also be understood in terms of the
Fock-state decomposition of the nucleus. The dominant Fock components are
determined by the resolution of the interaction, and a nucleus can resolve
more Fock states than a proton since the saturation scale $Q_s$ rises with
$A$. Thus, one can see that the leading parton distribution involves
higher multiparton Fock states in a nucleus and must fall more steeply
toward $x_F=1$, as suggested by the Blankenbecler-Brodsky counting rule
\cite{bb} (see also \cite{bf,bs}).

Note that this situation is analogous to what happens when one of the
bound nucleons in a relativistic nucleus is moving with a momentum higher
than the average momentum of an individual nucleon. This is the mechanism
of production of particles with $x_F>1$ in nuclear collisions \cite{bs}.

The involvement of higher Fock states means that gluon bremsstrahlung is
more intense in the interaction on a nucleus than on a proton target, i.e.
leads to larger energy loss.  Because of this, the large-$x_F$ suppression
may be envisioned to be a consequence of induced energy loss. Remarkably,
such an induced energy loss proportional to energy results in $x_F$
scaling. This is different from the energy-independent mean rate of energy
loss found in \cite{n,bh,bdms}. The latter, however, was calculated with
no restriction on the gluon radiation spectrum, a procedure that is not
appropriate when $x_F\to 1$ \cite{knp,knph}. 

The nuclear effects under discussion violate QCD factorization. This
happens because any hard reaction is a LRG process in the limit $x_F\to
1$, e.g. gluon radiation, is forbidden by energy conservation throughout a
rather large rapidity interval. On the other hand, factorization relates
this process to the parton distribution functions measured in other
reactions, for instance deep-inelastic scattering (DIS), which do not have
such a restriction. The light-cone dipole description employed here does
not involve a twist decomposition, but apparently the lack of
factorization does not disappear with increasing scale. This will be
confirmed below by explicit calculations for different hard reactions.

Thus, we conclude that the nuclear suppression imposed by energy
conservation as $x_F\to 1$ is a leading twist effect, breaking QCD
factorization. This is at variance with the conclusion of \cite{hoyer},
made on the basis of observed violation of $x_2$ scaling in $J/\Psi$
production off nuclei and assuming validity of the factorization theorem
\cite{factorization} even in the limit $x_F\to 1$.

Note that factorization is broken even for reactions on a free proton. For
instance, the Drell-Yan reaction, which is a LRG process as $x_F\to 1$,
cannot be expressed in terms of hadronic structure functions measured in
deep-inelastic scattering.

\section{Disappearance of nuclear shadowing at smallest
\boldmath$x_2$}\label{shadowing}

It is convenient to study shadowing and other coherence effects in the
rest frame of the nucleus. In this frame, the parameter controlling the
interference between amplitudes of the hard reaction occurring on
different nucleons is the longitudinal momentum transfer, $q_L$, related
to the coherence length, $l_c=1/q_L$. The condition for the appearance of
shadowing in a hard reaction is the presence of a coherence length that is
long compared to the nuclear radius, $l_c\gsim R_A$.

Clearly at large $x_1\sim 1$, a hard reaction is mediated by a projectile
valence quark and looks like a hard excitation of the valence quark in
this reference frame. For instance, the Drell-Yan (DY) process looks like
radiation of a heavy photon/dilepton by a valence quark \cite{hir}. The
coherence length in this case is related to the mean lifetime of a
fluctuation $q\to q\bar ll$ and reads \cite{hir,e-loss},
 \beq
l_c=\frac{2E_q\,\alpha(1-\alpha)}
{k_T^2+(1-\alpha)M_{\bar ll}^2+\alpha^2\,m_q^2}\ .
\label{10}
 \eeq
 Here $\vec k_T$ and $\alpha$ are the transverse momentum and the fraction
of the light-cone momentum of the quark carried by the dilepton; $M_{\bar
ll}$ is the effective mass of the dilepton; and $E_q=x_q\,s/2m_N$ and
$m_q$ are the energy and mass of the projectile valence quark. This simple
kinematic formula reflects the relation between the longitudinal momentum
transfer $q_L=(M_{q\bar ll}^2-m_q^2)/2E_q$ and the coherence length
$l_c=1/q_L$. Note that the fraction of the proton momentum $x_q$ carried
by the valence quark in this reference frame is not equal to $x_1$, but
$\alpha x_q=x_1$.  Clearly, when $x_1\to 1$, also $\alpha\to 1$, i.e. the
coherence length Eq.~(\ref{10}) vanishes in this limit, and no shadowing
is possible.

The onset of shadowing as function of rising coherence length can be
approximated with good accuracy as,
 \beq
R^{shad}_{A/N} \approx
1\,-\, {1\over4}\,\sigma_{eff}\,\la T\ra\,F_A^2(q_L)\ ,
\label{12}
 \eeq
 Here $R_{A/N}$ is the normalized ratio of the reaction rates on nuclear
and nucleon targets. $F_A(q_c)$ may be called the longitudinal nuclear
form factor, defined as
 \beq F_A^2(q_L) = \frac{1}{\la T_A \ra}\,
\int d^2b\, \left|\int\limits_{-\infty}^{\infty} dz\,
e^{iz/l_c}\,\rho_A(b,z)\right|^2,
 \label{14}
 \eeq
 where $\la T_A \ra=1/A\int d^2b\,T^2_A(b)$ is the nuclear thickness
$T_A(b)$ averaged over impact parameter $b$ and is evaluated as the
integral of the nuclear density over the longitudinal coordinate,
$T_A(b)=\int dz\,\rho_A(b,z)$. The effective cross section
 \beq
\sigma_{eff}(x_1,x_2,s)=\frac{\left\la
\sigma^2_{\bar qq}(\alpha r_T)\right\ra}
{\left\la
\sigma_{\bar qq}(\alpha r_T)\right\ra}\ ,
\label{13}
\eeq
 was evaluated in \cite{e-loss}.

\begin{figure}[tbh]
\includegraphics{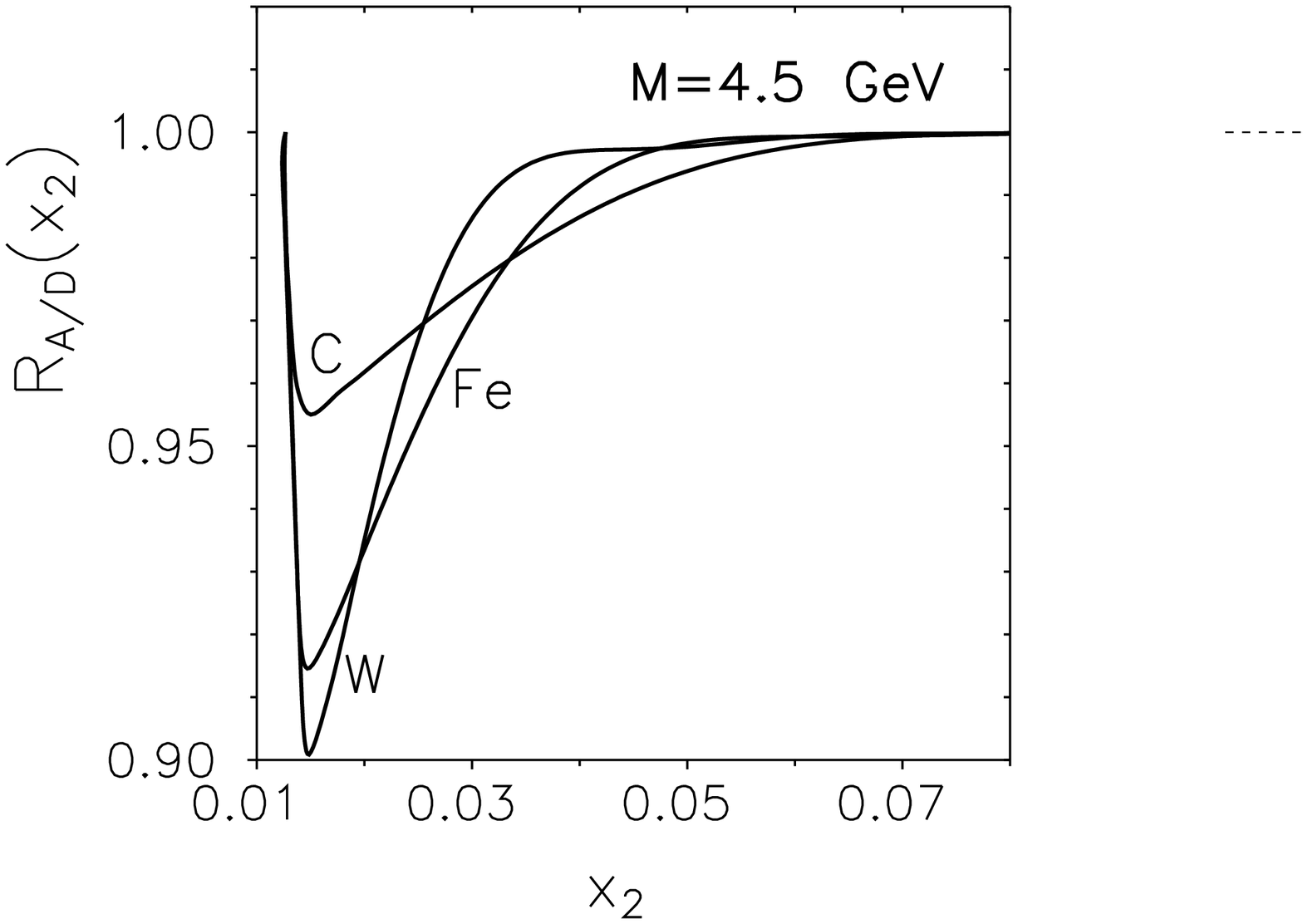}
\vspace{6.5cm}
{\caption[Delta]
 {Shadowing in DY reaction on carbon, iron and tungsten as function of
$x_2$ at $M_{\bar ll}=4.5\,GeV$ and $s=1600\GeV^2$. Nuclear shadowing
disappears both at large and small $x_2$, since the coherence length,
Eq.~(\ref{10}), vanishes in both limits.}
 \label{shad-4.5}}
 \end{figure}

With the mean value of $l_c$ given by Eq.~(\ref{14}), the
nucleus-to-deuteron ratios of DY cross sections are presented for
different nuclei in Fig.~\ref{shad-4.5} \cite{e-loss}.

 As expected, in accordance with Eq.~(\ref{10}), shadowing vanishes at the
smallest value of $x_2$ that can be accessed in DY reactions, while
according to QCD factorization it is expected to reach maximal strength.

 Note that the shrinkage of the coherence length towards $x_F=1$ does not
only lead to the disappearance of shadowing in this kinematic limit, but
also reduces shadowing in DY compared to DIS throughout the entire range
of $x_2$.

Note that the coherence length Eq.~(\ref{10}) linearly rises with energy,
therefore the interval of $x_1$, where the coherence length contracts down
to the nuclear size, shrinks like $1/E$ (see Sect.~\ref{gl-sh}).

\section{Sudakov suppression, or survival probability of large rapidity
gaps}\label{sudakov}

The BRAHMS experiment \cite{brahms} found a substantial nuclear
suppression for negative hadrons produced with high $p_T$ at large
pseudorapidity $\eta=3.2$, instead of the usual enhancement (see
Fig.~\ref{brahms}). Since the data cover rather small $x_2\sim 10^{-3}$,
it is tempting to interpret the suppression as either a result of
saturation \cite{glr,al} or the color glass condensate \cite{mv}, expected
in some models \cite{kkt}.

Note, however, that the data span a region of rather large $x_F$, where
all known reactions, both hard and soft, show considerable nuclear
suppression. Moreover, available data indicate that this effect scales
with $x_F$, rather than with $x_2$, as one would expect if the scaling
were the net effect of coherence.

Indeed, the collection \cite{barton,geist,data} of data depicted in
Fig.~\ref{geist} for the production of different species of hadrons in
$pA$ collisions at different energies, with the nuclear dependence
parametrized as $A^\alpha$, exhibit universality and $x_F$ scaling.

Data from the E866 experiment at Fermilab for nuclear effects in $J/\Psi$
and $\Psi^\prime$ production \cite{e866-psi}, shown in Fig.~\ref{psi},
also exhibit a strong suppression that is seen to scale in $x_F$ compared
to lower-energy data \cite{na3} and that appears universal when compared
with $\Psi^\prime$ \cite{e866-psi}. Recent measurements of nuclear effects
for $J/\Psi$ production in $D-Au$ collisions by the PHENIX collaboration
\cite{psi-qm} at RHIC are consistent with $x_F$ scaling, but they exhibit
a dramatic violation of $x_2$ scaling when compared with the $E866$ data
\cite{e866-psi}.

The DY reaction is also known to be considerably suppressed at large $x_F$
\cite{e-loss}, as one can see in Fig.~\ref{dy}. Unfortunately, no data
sufficiently accurate to test $x_F$ scaling are available at other
energies.

There is a feature common to all these reactions; namely, when the final
particle is produced with $x_F\to 1$, insufficient energy is left to
produce anything else. As a class, such events are usually called large
rapidity gap processes. Obviously, the restriction of energy conservation
may cause substantial suppression.  This is analogous to what happens in
QED when elastic electron scattering occurs with no bremsstrahlung within
a given resolution; it is described by what is known as the Sudakov form
factor.  The LRG cross section is more strongly suppressed as the
resolution improves.

If a large-$x_F$ particle is produced, the rapidity interval to be kept
empty is $\Delta y=-\ln(1-x_F)$. We describe particle production via
perturbative gluon radiation \cite{gb} with subsequent nonperturbative
hadronization.  Assuming as usual an uncorrelated Poisson distribution for
gluons, the Sudakov suppression factor, i.e. the probability to have a
rapidity gap $\Delta y$, becomes
 \beq
S(\Delta y) = e^{-\la n_G(\Delta y)\ra}\ ,
\label{20}
 \eeq
 where in our case $n_G(\Delta y)$ is the mean number of gluons that would
be radiated within $\Delta y$ if energy conservation were not an issue.

Note that even in the case where no gluon is radiated within the rapidity
gap, the hadronization can easily fill the gap with particles. The
probability that this does not happen is another suppression factor which,
however, is independent of target and cancels in the nucleus-to-proton
ratio.

The mean number $\la n_G(\Delta y)\ra$ of gluons radiated in the rapidity
interval $\Delta y$ is related to the height of the plateau in the gluon
spectrum, $\la n_G(\Delta y)\ra=\Delta y\,dn_G/dy$. Then, the Sudakov
factor acquires the simple form,
 \beq
S(x_F) = (1-x_F)^{dn_G/dy}\ .
\label{40}
 \eeq
 The height of the gluon plateau was estimated by Gunion and Bertsch
\cite{gb} as,
 \beq
\frac{dn_G}{dy} = \frac{3\alpha_s}{\pi}\,
\ln\left(\frac{m_\rho^2}{\Lambda_{QCD}^2}\right)\ .
\label{50}
 \eeq
 The value of $\alpha_s$ was fitted \cite{gb} to data on pion multiplicity
in $e^+e^-$ annihilation, where it was found that $\alpha_s=0.45$. This is
close to the critical value $\alpha_s=\alpha_c=0.43$ \cite{gribov} and to
the value $\la \alpha_s\ra=0.38$ calculated within a model of small
gluonic spots when averaged over the gluon radiation spectrum \cite{k3p}.
For further calculations we take $\alpha_s=0.4$, which gives with high
accuracy $dn_G/dy=1$, i.e. the Sudakov factor,
 \beq
S(x_F)=1-x_F\ .
\label{60}
 \eeq
Amazingly, this coincides with the suppression factor applied to
every additional Pomeron exchange in the quark-gluon string
\cite{kaidalov} and dual parton \cite{capella} models based on the
Regge approach.

Clearly, on a nuclear target, the Sudakov suppression factor should fall
more steeply as $x_F\to 1$ since multiple interactions enhance the
transverse kick given to the projectile parton and therefore tend to shake
off (i.e.  to radiate) more gluons. This can be understood in terms of the
Fock state decomposition. Specifically, according to the counting rules
\cite{bb}, the behavior of the single parton distribution function for
$x_1\to 1$ depends on the number of constituents in the particular Fock
state. A nucleus having a higher resolution, controlled by the saturation
scale $Q_s$ \cite{mv,al}, resolves more constituents and thus results in a
steeper fall off of the distribution function toward $x_1=1$.

We come to the nontrivial conclusion that the effective parton
distribution function in the beam hadron depends on the target. Such a
process-dependence constitutes an apparent breakdown of QCD factorization
and is a leading twist effect.

One can also formulate this suppression as $x_F\to 1$ as a survival
probability of the LRG in multiple interactions with the nucleus. Clearly,
every additional inelastic interaction contributes an extra suppression
factor $S(x_F)$. The probability of an n-fold inelastic collision is
related to the Glauber model coefficients via the
Abramovsky-Gribov-Kancheli (AGK)  cutting rules \cite{agk}.
Correspondingly, the survival probability at impact parameter $\vec b$
reads,
 \beqn
W^{hA}_{LRG}(b) &=&
\exp[-\sigma_{in}^{hN}\,T_A(b)]
\nonumber\\ &\times&
\sum\limits_{n=1}^A\frac{1}{n!}\,
\left[\sigma_{in}^{hN}\,T_A(b)\,
S^n(x_F)\right]^n\ .
 \label{70}
 \eeqn
 In this expression particles (gluons) are assume to be produced
independently in multiple rescattering, i.e. in Bethe-Heitler regime. Of
course it should be corrected for effects of coherence which turn out to
be either small or absent, i.e. can be neglected. Indeed, at small $x_F$
the Sudakov factor $S(x_F\to0)\to 1$, and Eq.~(\ref{70}) takes the form of
the standard Glauber expression for absorptive hadron-nucleus cross
section. In this case the coherence effects are known as Gribov inelastic
shadowing corrections which are known to be quite small, few percent 
\cite{mine}.

In another limiting case $x_F\to 1$ energy conservation allows only
radiation of low-energy gluons having short coherence time. Therefore,
particles are produced incoherently in multiple interactions, and
Eq.~(\ref{70}) is legitimate.

At large $x_F\sim 1$, Eq.~(\ref{70}) is dominated by the first term;
therefore, integrating over impact parameter, one gets for the
nucleus-to-proton ratio, $R_{A/p}(x_F\to1)\sim A^{1/3}$. This expectation
is confirmed by a measurement \cite{helios} of the $A$-dependence of the
cross section for the LRG process $pA\to pX$, quasi-free diffractive
excitation of the nucleus. The single diffraction cross section was found
to be
 \beq
\sigma^{pA}_{diff} = \int\limits_{0.925}^1
dx_F\,\frac{d\sigma(pA\to pX)}{dx_F}=
\sigma_0\,A^\alpha,
\label{80}
 \eeq
 with $\alpha=0.34\pm0.02$, consistent with the above expectation.

\section{Production of leading hadrons with small
\boldmath$p_T$}\label{small-pt}

\begin{figure}[tbh]
\includegraphics{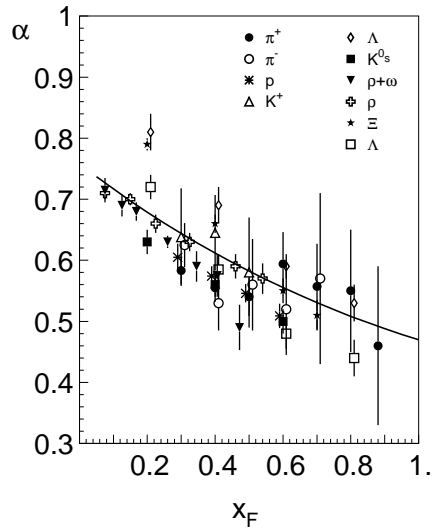}
\vspace{7.5cm}
{\caption[Delta]
 {The exponent describing the $A$-dependence ($\propto A^\alpha$) of the
ratio for the production of different hadrons in $p-Au$ relative to $pp$
collisions as function of $x_F$. The collection of data and references can
be found in \cite{barton,geist,data}. The curve is a result of a parameter
free calculation using Eq.~(\ref{120}).}
 \label{geist}}
 \end{figure}

The collection of data from \cite{barton,geist,data} for the production of
different species of particles in $pA$ collisions, depicted in
Fig.~\ref{geist}, exhibits quite strong and universal nuclear suppression
at large $x_F$. Moreover, these data spanning the lab energy range from 70
to 400 GeV, demonstrate that the nuclear effects scale in $x_F$.

It is natural to relate the observed suppression to the dynamics discussed
above, which is close to the description of soft inclusive reactions
within the quark-gluon string \cite{kaidalov}, or dual parton
\cite{capella} models.

 The nuclear effects can be calculated summing oven $n$ and integrating
over impact parameter in Eq.~(\ref{70}),
 \beqn
 R_{A/N}(x_F) &=&
\frac{1}{(1-x_F)\sigma_{abs}A}
\int d^2b\,e^{-\sigma_{abs}T_A(b)}\,
\nonumber\\ &\times&
\left[e^{(1-x_F)\sigma_{abs}\,T_A(b)}
-1\right].
\label{120}
 \eeqn
 In the Glauber model, the effective cross section is the familiar
inelastic $NN$ scattering cross section.  However, the actual number of
collisions, which determines the value of the effective absorption cross
section, is subject to considerable modification from Gribov's inelastic
shadowing corrections, which make the nuclear medium substantially more
transparent.  These corrections considerably reduce both the number of
collisions and the effective absorption cross section $\sigma_{abs}$
\cite{mine,kps}.

In order to compare with data, the nuclear effects are parametrized as
$R_{A/N}\propto A^\alpha$, where $\alpha$ varies with $A$. We use A=40,
for which the Gribov corrections evaluated in \cite{kps} lead to an
effective absorption cross section $\sigma_{abs}\approx 20\mb$. Then the
simple expression Eq.~(\ref{120}) explains the observed $x_F$ scaling and
describes rather well the data depicted in Fig.~\ref{geist}.

One may wonder why $\alpha(x_F)$ plotted in Fig.~\ref{geist} does not
reach values as small as $1/3$, even when $x_F\to 1$. As mentioned, this
exponent varies with $A$, and simple geometrical considerations may be
accurate only for sufficiently heavy nuclei. In the case of diffraction
\cite{kps}, there is a specific enhancement of the effective absorption
cross section that makes $\alpha$ smaller in this case, in good accord
with data \cite{helios}.

Note that our description is very close to that in the dual parton model
(or quark-gluon string model) \cite{bckt}. However, we present a different
interpretation of the same phenomena and introduce Gribov corrections for
inelastic shadowing, which substantially reduce the number of collisions.

\section{High-\boldmath$p_T$ hadron production at forward rapidities,
the BRAHMS data}\label{large-pt}

\begin{figure}[tbh]
\includegraphics{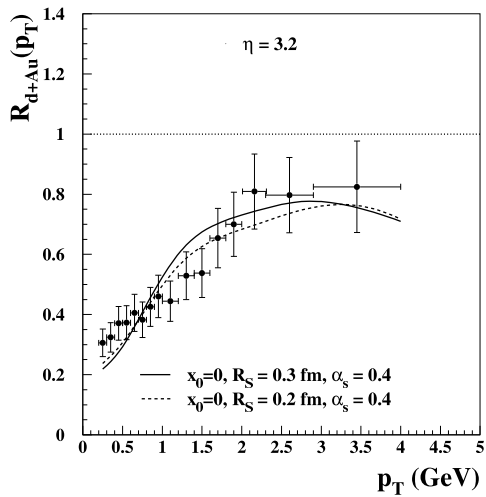}
\vspace{7cm}
{\caption[Delta]
 {Ratio of negative particle production rates in $d-Au$ and $pp$
collisions as function of $p_T$. Data are from \cite{brahms}, solid and
dashed curves correspond to calculations with the diquark size $0.3\fm$
and $0.4\fm$ respectively. }
 \label{brahms}}
 \end{figure}

The BRAHMS collaboration performed measurements of nuclear effects for
production of negative hadrons at pseudorapidity $\eta=3.2$ and transverse
momentum up to $p_T\approx 4\GeV$. Instead of the usual Cronin enhancement
a suppression was found, as one can see from Fig.~\ref{brahms}.

First, consider the rather strong suppression of the data at small $p_T$.
One can understand this in terms of the simple relation for the
$p_T$-integrated cross sections,
 \beq
\int d^2p_T\,\frac{d\sigma}{d\eta d^2p_T}\,
=\left\la\frac{dn}{d\eta}\right\ra\,
\sigma_{in}\ .
\label{85}
 \eeq
 The mean number of produced particles per unit rapidity $\la dn/d\eta\ra$
has an $A$-dependence which varies with rapidity. Particle production at
mean rapidity is related to the radiation of gluons, whose multiplicity
rises as $dn_G/d\eta\propto A^{1/3}$ (for the moment we neglect gluon
shadowing and assume the Bethe-Heitler regime for gluon radiation). Since
the inelastic cross section $\sigma^{pA}_{in}\propto A^{2/3}$, the
integrated inclusive cross section, Eq.~(\ref{85}), rises linearly with
$A$. This is in accordance with the AGK cancellation \cite{agk} of
shadowing for the inclusive cross section known as the Kancheli-Mueller
theorem.

Nuclei modify the $p_T$ distribution of radiated gluons, an effect known
as the color glass condensate (CGC) \cite{mv,al} or Cronin effect
\cite{cronin,knst}. Due to this effect, gluons are suppressed at small
$p_T$, enhanced at medium $p_T$, and are unchanged at large $p_T$. Gluon
shadowing, or the Landau-Pomeranchuk effect, is a part of the CGC and
reduces the total number of radiated gluons more strongly at small than at
large $p_T$. Thus, the observed strong suppression of small $p_T$ particle
production at mid rapidities is a manifestation of the CGC.

One, however, should be careful with the interpretation of data in terms
of the CGC, which is supposed to be a result of coherence between
different parts of the nucleus. It turns out that nuclear modifications of
the transverse momentum distribution occur both in the coherent and
incoherent regimes.  While the former can be an effect of the CGC, the
latter has little to do with this phenomenon. In particular, the RHIC data
at mid rapidities are in the transition region, i.e. particles are
produced coherently on the nucleus at small $p_T\lsim 1\GeV$, but
incoherently at larger $p_T$ \cite{knst}.

The suppression at small-$p_T$ observed at $\eta=3.2$ is even stronger
than at mid rapidities.  At this rapidity, the overall scale of the
suppression is related to the fact that particle production is dominated
by fragmentation of the projectile valence quarks. Gluons are additionally
suppressed due to softness of the gluon fragmentation function leading to
a substantially larger value of $x_1$ for gluons than for pions.  
Therefore, the origin of the the suppression is quite different from that
at mid rapidity \cite{mytalk}.  Because the number of valence quarks is
fixed and equal to three when integrated over rapidity (Gottfried sum
rule), the number of valence quarks produced with $x_F\to 1$ must be even
smaller, and accordingly the ratio of the $p_T$-integrated inclusive cross
sections should be suppressed well below unity. In this case we can use
either our results or the data plotted in Fig.~\ref{geist}, both of which
suggest a suppression factor of approximately $A^{-0.3}\approx 0.2$, in
good agreement with the BRAHMS data. This suppression is not affected
either by the CGC or gluon shadowing.

Note that the dominance of valence quarks in the projectile proton leads
to an isospin-biased ratio. Namely, negative hadrons with large $p_T$
close to the kinematic limit are produced mainly from $u$, rather than
$d$, quarks.  Therefore, more negative hadrons are produced by deuterons
than by protons, and this causes an enhancement of the ratio plotted in
Fig.~\ref{brahms} by a factor of $3/2$ \cite{mytalk}. Further on, we take
care of this by using proper fragmentation functions for negative hadrons.

The cross section of hadron production in $dA(pp)$ collisions is given by
a convolution of the distribution function for the projectile valence
quark with the quark scattering cross section and the fragmentation
function,
 \beqn
&&\frac{d\sigma}{d^2p_T\,d\eta} =
\sum\limits_q \int\limits_{z_{min}}^1 dz\,
f_{q/d(p)}(x_1,q_T^2)
\nonumber\\ &\times&
\left.\frac{d\sigma[qA(p)]}{d^2q_T\,d\eta}
\right|_{\vec q_T=\vec p_T/z}\,
D_{h^-/q}(z)
\label{87}
 \eeqn
 Here
 \beq
x_1=\frac{q_T}{\sqrt{s}}\
e^\eta\ .
 \label{88}
 \eeq

 We use the LO GRV parametrization \cite{grv} for the quark distribution
in the nucleon. As we explained above, the interaction with a nuclear
target does not obey factorization, since the effective projectile quark
distribution correlates with the target. Summed over multiple
interactions, the quark distribution in the nucleon reads,
 \beqn
&&f^{(A)}_{q/N}(x_1,q_T^2)=
C\,f_{q/N}(x_1,q_T^2)\,
\nonumber\\ &\times&
\frac{\int d^2b\,
\left[e^{-x_1\sigma_{eff}T_A(b)}-
e^{-\sigma_{eff}T_A(b)}\right]}
{(1-x_1)\int d^2b\,\left[1-
e^{-\sigma_{eff}T_A(b)}\right]}\ .
 \label{89}
 \eeqn
 Here the normalization factor $C$ is fixed by the Gottfried sum rule.

The cross section of quark scattering on the target in Eq.~(\ref{87}) is
calculated in the light-cone dipole approach \cite{zkl,jkt}, which
provides an easy way to incorporate multiple interactions. Obviously, the
$p_T$ distribution of hadrons in the final state is affected by the
primordial transverse motion of the projectile quarks.  In our
calculation, we separate the contributions characterized by different
initial transverse momenta and sum over three different mechanisms of
high-$p_T$ production.

\subsection{Quark-diquark break up of the proton.}

We employ the quark-diquark model of the proton with the $\widehat{ud}$
diquark small compared to the proton radius \cite{bs,review}.
Correspondingly, the third valence quark external to the diquark has much
smaller transverse momentum than the two others sitting inside the
diquark.  As a first mechanism, we consider proton breakup $p\to \widehat
{qq}+q$. We treat the diquark $\{qq\}$ as point-like and integrate over
its momentum. Then the $k_T$ distribution of the projectile valence quark,
after propagation through the nucleus at impact parameter $\vec b$, is
given by \cite{kst1,kst2},
 \beqn
&&\frac{d\sigma(NA\to qX)}{d^2k_T\,d^2b} =
\int \frac{d^2r_1 d^2r_2}{(2\pi)^2}\
e^{i\vec k_T(\vec r_1-\vec r_2)}
\nonumber\\ &\times&
\Psi^\dagger_N(r_1)\,\Psi_N(r_2)
\left[1+e^{-{1\over2}\sq(\vec r_1-\vec r_2)T_A(b)}
\right. \nonumber\\ &-& \left.
e^{-{1\over2}\sq(\vec r_1)T_A(b)} -
e^{-{1\over2}\sq(\vec r_2)T_A(b)}
\right]\ .
\label{90}
 \eeqn
 Here the quark-diquark wave function of the nucleon is taken in a form
that matches the known perturbative QCD behavior at large transverse
momenta, $\Psi_N(r)\propto K_0(r/R_p)$, where $K_0$ is the modified Bessel
function, $R_p^2={4\over3}\, r_{ch}^2$, and $r_{ch}$ is the mean charge
radius of the proton. We assume that the quark's longitudinal momentum
dependence factorizes and is included in $f_{q/N}(x_1)$. The dipole cross
section $\sq(r)$ is taken in the saturated form \cite{kst2}, inspired by
the popular parametrization of Ref.~\cite{gbw}, but adjusted to the
description of soft data.

 This contribution dominates the low transverse momentum region $k_T \lsim
1\GeV$.

\subsection{Diquark break up \boldmath$\widehat{qq}\to qq$.}

At larger $k_T$ the interaction resolves the diquark, so its break-up
should be considered. This contribution is calculated in accordance with
Refs.~\cite{kst1,kst2},
 \beqn
&&\frac{d\sigma(\widehat{qq}A\to qX)}{d^2k_T\,d^2b} =
\int \frac{d^2r_1 d^2r_2}{2\,(2\pi)^2}\
e^{i\vec k_T(\vec r_1-\vec r_2)}
\nonumber\\ &\times&
\Psi^\dagger_D(r_1)\,\Psi_D(r_2)
\left[2-e^{-{1\over2}\sq(\vec r_1)T_A(b)}
\right. \nonumber\\ &-&
e^{-{1\over2}\sq(\vec r_2)T_A(b)}-
e^{-{1\over2}\sq(\vec r_2/2)T_A(b)}-
\nonumber\\ &-&
e^{-{1\over2}\sq(\vec r_2/2)T_A(b)}-
e^{-{1\over2}\sq(\vec r_1-{1\over2}\vec r_2)}
\nonumber \\ &-&
e^{-{1\over2}\sq(\vec r_2-{1\over2}\vec r_1)}+
2\,e^{-{1\over2}\sq(\vec r_1-\vec r_2)T_A(b)}
\nonumber \\ &+& \left.
2\,e^{-{1\over2}\sq(\frac{\vec r_1-\vec r_2}{2})T_A(b)}
\right]\ .
\label{100}
 \eeqn
 The diquark wave function is also assumed to have a Bessel function form,
with a mean quark separation of $0.2-0.3\fm$. There is much evidence that
such a small diquark represents the dominant quark configuration in the
proton \cite{review}.

\subsection{Hard gluon radiation \boldmath$q\to Gq$.}\label{gluons}

At large $k_T$, the dipole approach should recover the parton model
\cite{3f}, where high momentum transfer processes occur (in leading order)
as binary collisions with the transverse momentum of each final parton of
order $k_T$.  Clearly, this is different from the description in
Eqs.~(\ref{90})-(\ref{100}), where one assumes that the projectile valence
quark acquires high transverse momentum as a result of multiple
rescatterings, while the radiated gluons that balance this momentum are
summed to build up the dipole cross section. The latter is fitted to DIS
data involving gluons of rather low transverse momenta. Therefore, one
should explicitly include in the dipole description radiation of a gluon
with large transverse momentum that approximately balances $k_T$, i.e. the
process $qN\to qGX$. In the dipole approach, the cross section is given by
the same formula, Eq.~(\ref{90}), except that the nucleon wave function is
replaced by the quark-gluon light-cone wave function, $\Psi_N(r_T)
\Rightarrow \Psi_{qG}(r_T)$ \cite{km,kst1}, where \cite{kst2}
 \beq
\Psi_{qG}(\vec r_T)= -\frac{2i}{\pi}\,
\sqrt{\frac{\alpha_s}{3}}\
\frac{\vec r_T\cdot\vec e^{\,*}}{r_T^2}\,
{\rm exp}\left(-{r_T^2\over2r_0^2}\right)\ .
\label{110}
 \eeq
 and $r_0=0.3\fm$. Such a small mean quark-gluon separation is a result of
a phenomenological analysis of data for soft single diffraction $pp\to
pX$. The only way to explain the abnormally small triple-Pomeron coupling,
which is translated into very weak diffractive gluon radiation, is to
assume that the Weizs\"acker-Williams gluons in the proton are located
within small spots \cite{kp}. The spot size $r_0$ was fitted to
diffractive data \cite{kst2}, and the result $r_0=0.3\fm$ agrees with both
lattice calculations \cite{pisa} and also with the phenomenological model
of the instanton liquid \cite{shuryak}.

Notice that at small $k_T$ there is a risk of double counting in such a
procedure, since the radiated gluon may be counted twice, explicitly and
implicitly as a part of the dipole cross section. However, the first two
contributions, Eqs.~(\ref{90})-(\ref{100}), and the last one dominate in
different regions of $k_T$ and we found their overlap very small.

\subsection{Gluon shadowing}\label{gl-sh}

Although the BRAHMS data involve rather large values of $x_F$, the
corresponding values of $x_2$ are so small that the considerations of
Sects.~\ref{shadowing} and \ref{psi-sect} lead to little reduction of the
coherence length. Nevertheless, gluon shadowing corrections are expected
to be quite small even at very small $x_2$ (e.g. see predictions for LHC
in \cite{knst}). This is related to the presence of small size gluonic
spots in nucleons, discussed in the previous Sect.~\ref{gluons}. Indeed,
one goes to small $x_2$ in order to make the coherence length longer
($\l_c\propto 1/x_2$) and thus to arrange an overlap between the parton
clouds which belong to different nucleons separated in the longitudinal
direction. However, even if this condition is fulfilled, the gluon clouds
hardly overlap in impact parameter, if they are shaped as small size
spots. Expecting the strongest gluon shadowing at $Q^2\to 0$ one can
estimate the shadowing correction to be,
 \beq
R_G(b)\sim 1-\exp\left[-\frac{3\pi}{4}\,r_0^2\,T_A(b)\right]\ .
\label{115}
 \eeq
 For heavy nuclei this estimate gives a rather weak shadowing, about
$10\%$, in good accord with more accurate calculations \cite{kst2,krt2} or
with the NLO analysis of nuclear DIS data \cite{florian}.

We calculate gluon shadowing corrections within the dipole approach with
the light-cone gluon distribution function, Eq.~(\ref{110}). The details
can be found in \cite{kst2,krt2}.

\subsection{Comparison with data}\label{data}

\begin{figure}[tbh]
\includegraphics{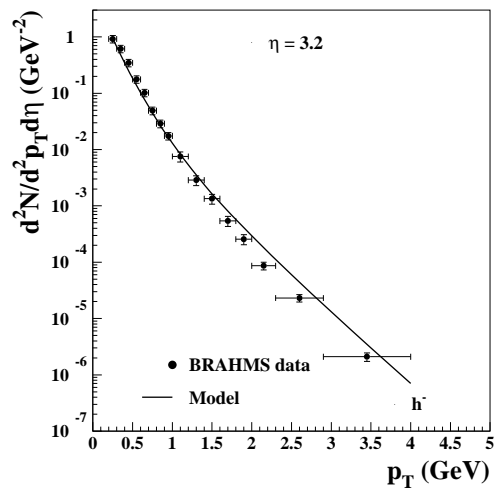}
\vspace{7cm}
{\caption[Delta]
 {Number of negative hadrons versus $p_T$ produced in $pp$ collisions at
$\sqrt{s}=200\GeV$ and pseudorapidity $\eta=3.2$. Our
calculations, given by the solid curve, are compared with BRAHMS
data \cite{brahms}.}
 \label{pp}}
 \end{figure}

First of all, one should confront the model with the
$p_T$-dependent cross section of hadron production in $pp$
collisions. Although the nuclear effects under discussion are not
sensitive to this dependence, which mostly cancels in the $dA/pp$
ratio, this would be a stringent test of the model. Our
calculations are compared with $pp$ data from the BRAHMS
experiment at $\eta=3.2$ in Fig.~\ref{pp}.
 In view of the isospin asymmetry of leading particle production mentioned
above, it is important to use proper fragmentation functions. The
standard ones extracted from data on $e^+e^-$ annihilation give a
sum of positive and negative hadrons, while one needs only
fragmentation functions for production of negative hadrons in
order to compare with the BRAHMS data. We use these negative
fragmentation functions in our calculations \cite{hermes}.

Now we are in a position to predict nuclear effects employing the dipole
formalism and the mechanism described above. The results are compared with
the BRAHMS data for the minimum-bias ratio\cite{brahms} in
Fig.~\ref{brahms}. One can see that this parameter-free calculation does
not leave much room for other mechanisms, including a strong CGC. On the
other hand, our calculations do include the CGC via explicit gluon
radiation and via gluon shadowing. This is, however, a rather moderate
effect due to smallness of the gluonic spots in nucleons
\cite{kst2,kp,shuryak}, as is described by the quark-gluon wave function
in Eq.~(\ref{110}).

\begin{figure}[tbh]
\includegraphics{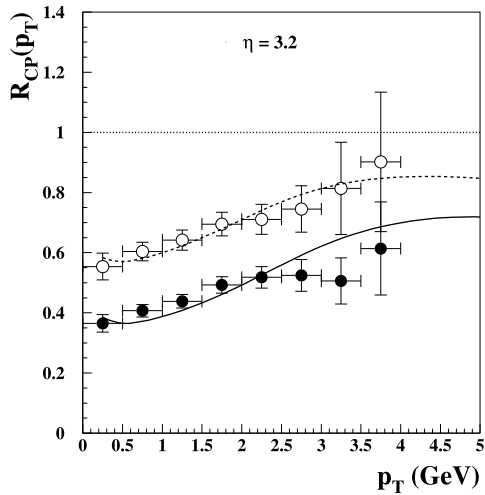}
\vspace{7cm}
{\caption[Delta]
 {Ratio of negative particle production in central (0-20\%) and
semi-central (30-50\%) to peripheral (60-80\%) $d-Au$ collisions, shown by
closed and open points respectively. The results of the corresponding
calculations are depicted by solid and dashed curves.}
 \label{centrality}}
 \end{figure}

It is interesting also to check whether the predicted dependence of the
ratio on impact parameter is supported by data. Our results for the ratio
of the cross sections of central and semicentral to peripheral collisions
are depicted in Fig.~\ref{centrality} in comparison with BRAHMS data
\cite{brahms}.
 The agreement is rather good.

Although the BRAHMS collaboration has presented their results for nuclear
effects as a function of rapidity, we skip this comparison, since it
cannot fail. Indeed, predictions \cite{knst} for the Cronin effect at
$\eta=0$ published in advance of data were quite successful. That region
is dominated by production and fragmentation of gluons. The very forward
region, which is under consideration now, is dominated by production and
fragmentation of valence quarks, and our calculations are successful here
as well.  For these reasons, we should not be much off the data at any
other (positive) rapidity, which differs only in the relative
contributions of valence quarks and gluons.

\section{Nuclear suppression of dileptons at large \boldmath$x_F$}

\begin{figure}[tbh]
\includegraphics{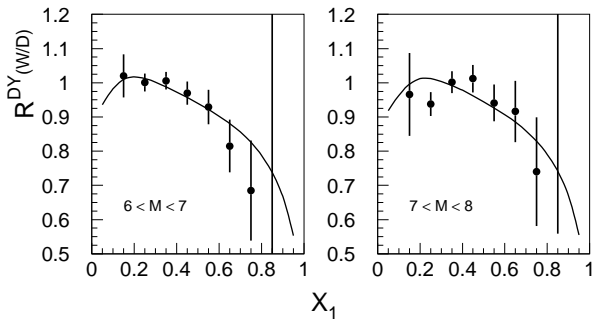}
\vspace{4.5cm}
{\caption[Delta]
 {Ratio of DY cross sections on tungsten and deuterium as a function of
$x_1$, at large dilepton masses to eliminate nuclear shadowing.}
 \label{dy}}
 \end{figure}

It was first observed in the E772 experiment at Fermilab \cite{e772} that
the DY process is suppressed at large $x_F$. Two mechanisms that may
possibly be responsible for this effect have been considered so far,
energy loss in the initial state \cite{kn1983,vasiliev,e-loss} and
shadowing \cite{e772,e-loss}. To make the interpretation more certain, and
to disentangle these two options, one can select data with the dilepton
effective mass sufficiently large to assure that the coherence length is
too short for shadowing. An example of such data for the ratio of tungsten
to deuterium is depicted in Fig.~\ref{dy}.

\subsection{ Rest Frame Description}

According to the rest frame interpretation \cite{hir} (see also
\cite{bhq,kst1,krt3,krtj,bms}), the DY process looks like fragmentation of
a projectile quark into a dilepton via bremsstrahlung of a heavy photon.
One can either calculate this perturbatively and use the quark
distribution function at the corresponding scale, or employ the soft quark
distribution functions and use a phenomenological fragmentation function
$q\to q\bar ll$. The latter approach, already used in \cite{e-loss}, is
appropriate here, and we will apply it using the cross sections of soft
production of valence quarks in $pp$ and $pA$ collisions fitted in the
following section. To the extent that these cross sections are subject to
nuclear suppression at large $x_F$, the DY process should be suppressed as
well.

We perform calculations within the same formalism used in \cite{e-loss},
but the source of suppression is not a simple initial state energy loss,
but an effective one that results from the nuclear modification of the
Fock state decomposition discussed above. The nucleus-to-deuterium ratio
reads,
 \beq
R_{A/D}(x_1)=
\frac{2\int\limits_{x_1}^1 dx\,
\frac{d\sigma(pA\to qX)}{dx}\
D_{\bar ll/q}({x_1\over x})}
{A\int\limits_{x_1}^1 dx\,
\frac{d\sigma(pD\to qX)}{dx}\
D_{\bar ll/q}({x_1\over x})}\ .
\label{140}
 \eeq Here we implicitly take into account the difference in the isospin
composition of deuterium and tungsten. We also incorporate the
contribution of projectile antiquarks and target quarks using the STEQ
quark distribution functions \cite{steq}.

\subsection{Nuclear suppression of valence quarks}

To evaluate Eq.~(\ref{140}), one needs to know the cross sections of soft
valence quark production in $pp$ and $pA$ collisions. To obtain this, we
turn the problem around, trying to be more model independent, and get the
nuclear suppression of valence quark jets directly from data. For this
purpose, we fitted data \cite{barton} for pion production in $pp$ and $pA$
collisions at $100\GeV$. We describe the spectrum of produced pions as,
 \beqn
&&\frac{d\sigma(pA\to\pi X)}{dx_1}
\nonumber\\ &=&
\int\limits_{x_1}^1 dx\,
\frac{d\sigma(pA\to qX)}{dx}\,
D_{\pi/q}(x/x_1)\ .
\label{130}
 \eeqn
 The fragmentation function $q\to\pi$, $D_{\pi/q}(z)$, is known. We use
the form suggested by the Regge approach \cite{kaidalov},
$D_{\pi^+/u}=D_{\pi^-/d}=(1-z)^{-\alpha_R+\lambda}$ and
$D_{\pi^+/d}=D_{\pi^-/u}=(1-z)D_{\pi^+/u}$, where $\lambda\approx1/2$.

\begin{figure}[tbh]
\includegraphics{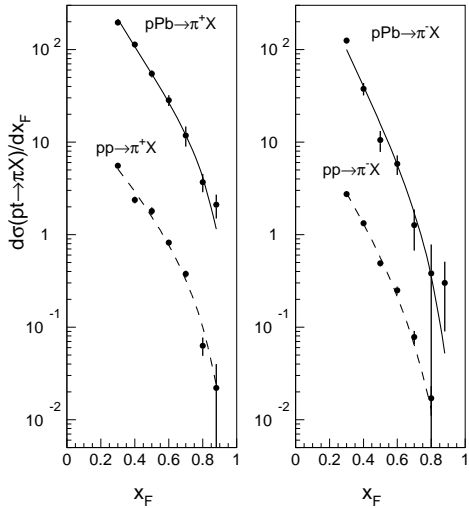}
\vspace{8cm}
{\caption[Delta]
 {Fit to data \cite{barton} for $pp(A)\to\pi^{\pm}$ with parametrization
described in text. The results are shown by dashed and solid curves for
$pp$ and and $p-Pb$ collisions respectively.}
 \label{fit}}
 \end{figure}

The unknown function in Eq.~(\ref{130}) is the cross section for quark jet
production in $pp$ or $pA$ collisions. We parametrized these cross
sections by a simple $x$-dependence, $\propto (1-x^\epsilon)^u/\sqrt{x}$
and $\propto (1-x^\epsilon)^d/\sqrt{x}$ for production of $u$ and $d$
quarks respectively, performed a fit to Fermilab data \cite{barton} at
$100\GeV$, and found $u=1.85\pm0.07$, $d=3.05\pm0.15$ for pp collisions;
$u=2.00\pm0.11$, $d=4.15\pm0.33$ for p-Al collisions; $u=2.03\pm0.14$,
$d=4.00\pm0.33$ for p-Pb collisions.  The quality of the fit can be seen
in Fig.~\ref{fit}.

While the parameter $u$ does not show any strong $A$-dependence, the value
of $d$ rises with $A$ making the $x$-dependence of the cross section
steeper. This is no surprise, since $d$ quarks have a steeper distribution
function in the proton. Unfortunately, the data are not sufficient to fix
well the parameter $\epsilon$, which is found to be $\epsilon=1.5\pm0.9$.

\subsection{Comparison with data}

Drell Yan results based on Eq.~(\ref{140}) are compared with the E772 data
in Fig.~\ref{dy}.  For $D_{\bar ll/q}(z,M^2)$ we used the phenomenological
fragmentation function $D_{\bar ll/q}(z,M^2)\propto (1-z)^{0.3}$ that was
fitted to data in Ref.~\cite{e-loss} and found to be independent of $M^2$
within errors, and for the quark production cross sections we used the fit
performed in the previous section. The parameter $\epsilon$, which is
poorly defined by the data, affects the $x_1$ dependence of DY cross
section ratio but hardly varies the amount of suppression at large $x_F$.
We select $\epsilon=1.3$, which is well within the errors and provides for
$R_{A/D}(x_1)$ a shape similar to the data.

Note that this description of nuclear effects appears different from the
energy-loss scenario employed in \cite{e-loss}. There is, however, a
strong overlap between the two mechanisms. The nuclear modification of the
projectile Fock decomposition is just a different way of calculating
nuclear effects due to gluon bremsstrahlung, which is the source of the
induced energy loss. Therefore the current approach might be called an
effective energy-loss description. The results, however, are different.
Namely, the simple mean energy loss of a single parton leads to nuclear
modifications that scale in $\Delta x_1$, i.e. vanish with increasing
energy at fixed $x_1$. In contrast, the current multi-parton effective
energy loss rises linearly with energy and leads to an $x_1$ scaling in
good accord with data.

Note that our interpretation of nuclear effects in the DY process is quite
different from the description in \cite{bckt}, where nuclear effects are
predicted to scale in $x_2$ and are possible only if the coherence length
is longer than the nuclear size. On the contrary, we expect the nuclear
suppression to scale in $x_F$.

\section{Charmonium suppression at large \boldmath$x_F$}\label{psi-sect}

Nuclear suppression of Charmonium production has been observed to be
steeply increasing at large $x_F\sim 1$ in many experiments. Understanding
this effect has been a challenge for a long time. Although the first data
\cite{na3} were well-explained and even predicted \cite{katsanevas} by an
energy-loss mechanism \cite{kn1983}, later data on $J/\Psi$ production at
higher energies demonstrated that this mechanism is not sufficient, since
it does not explain the observed $x_F$ scaling.

This problem was studied within the dipole approach in \cite{kth}. A
substantial part of the suppression was found to be a higher twist effect
related to the large size of charmonia. Such a suppression is frequently
identified with simple final-state absorption. However, at high energies a
$\bar cc$ fluctuation of a projectile gluon propagates and attenuates
through the entire nucleus \cite{kz1991,hk-lc,hikt2}; moreover, in the
case of hadroproduction, such a dipole is colored.  A description of this
process in terms of the light-cone dipole approach was developed in
\cite{kth}, and the cross section was calculated in a parameter-free way.

The rest of the suppression observed experimentally was prescribed to be
the effect of energy loss and gluon shadowing. Within this interpretation
of data, the observed $x_1$ scaling looked like an accidental compensation
of energy-loss corrections decreasing with energy and gluon shadowing
effects rising with energy. Correspondingly, an approximate $x_2$ scaling
(broken at low energies by energy loss) was predicted in \cite{kth} for
energies ranging between Fermilab (fixed targets) and RHIC. However, the
recent data from the PHENIX experiment at RHIC found a dramatic violation
of $x_2$ scaling in strict contradiction with this prediction. In fact,
any parton model based on QCD factorization predicts $x_2$ scaling and
contradicts this data.

We think that the higher twist nuclear shadowing was correctly calculated
in \cite{kth}, but that the gluon shadowing was miscalculated and led to
the incorrect predictions for RHIC \cite{kth,hkp}. What was missed in
\cite{kth} is the shrinkage of the coherence length towards the kinematic
limit.  This effect, found for the DY reaction in \cite{e-loss} and
discussed above in Sect.~\ref{shadowing}, is even more important for gluon
shadowing. Due to proximity of the kinematic limit for $J/\Psi$ production
with $x_F\to 1$, the effective value of $x_2$ is substantially increased,
 \beq
\tilde x_2 \sim \frac{x_2}{1-x_1}\ .
\label{150}
 \eeq
 Additionally, the coherence length available for gluon shadowing gets
another small factor $P^G$,
 \beq
l_c^G=\frac{P^G}{m_N\,\tilde x_2}
= \frac{s\,P^G}{M_{\Psi}^2\,m_N}\
x_1(1-x_1)\ .
\label{160}
 \eeq This factor was evaluated in \cite{krt2} as $P^G\approx 0.1$. Thus,
we conclude that for the kinematics of the E772/E866 experiments the
coherence length for gluon shadowing does not exceed $l_c^G\lsim 0.8\fm$,
i.e. no gluon shadowing is possible. Therefore, one should search for an
alternative explanation of the data.

\begin{figure}[tbh]
\includegraphics{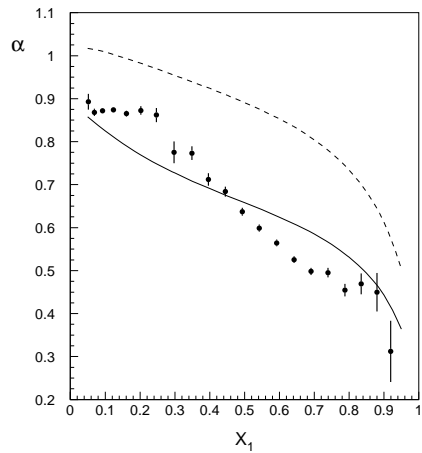}
\vspace{7cm}
{\caption[Delta]
 {Tungsten-to-beryllium cross section ratio for charmonium production as
function of $x_1$. Data are from the E866 experiment \cite{e866-psi}. The
dashed curve shows the contribution of the extra Sudakov suppression
extracted from data for soft hadron production. The solid curve also
includes the higher twist shadowing related to the nonzero $\bar cc$
separation.}
 \label{psi}}
 \end{figure}

Obviously, the same Sudakov effect that causes the large $x_F$ suppression
of other particles, in particular light hadrons and lepton pairs, affects
the charmonium production as well. We can use the same results for nuclear
softening of the produced valence quark, fitted to data, as for the DY
reaction. Nevertheless, the phenomenological fragmentation function
$q\to\Psi q$ should be different. Fitting data for $J/\Psi$ production in
$pp$ collisions we found $D_{\Psi/q}(z)\propto (1-z)^{1.6}$. Then, using
the same convolution of the distribution function of the produced quark
and the fragmentation function as in (\ref{140}), we arrive at the
suppression depicted by the dashed curve in Fig.~\ref{psi}.

 After the higher twist shadowing calculated in \cite{kth} is added, the
result is closer to the data, as shown by the solid curve in
Fig.~\ref{psi}. We see that the scale of the nuclear suppression at large
$x_F$ agrees with the data, although the shape of the $x_F$ dependence
needs to be improved. This problem is a consequence of the oversimplified
parametrization for the distribution function of the produced quark.
Unfortunately, the errors of the available data for light hadron
production are too large and do not allow use of more sophisticated
parametrizations.

Since the Sudakov suppression scales in $x_1$, one should expect an
approximate $x_F$ scaling, which is indeed observed in data. Therefore a
similar $x_F$ dependence is expected at RHIC and LHC, but the onset of
$x_2$ scaling, which has been naively expected at high energies
\cite{kth}, will never occur. Preliminary data from RHIC \cite{psi-qm}
confirm this. Additionally, they do not show any appreciable effect of
gluon shadowing, in spite of smallness of $x_2$. This agrees with the
finding of \cite{kst2}, that the leading twist gluon shadowing is very
weak. There is the possibility of a strong higher twist effect that may
make gluon shadowing in the charmonium channel stronger than in DIS
\cite{kth}. Such a prediction is based, however, on an ad hoc
phenomenological potential model and may be incorrect.

Open charm production is expected to have a similar Sudakov nuclear
suppression, like that shown by the dashed curve in Fig.~\ref{psi}. The
higher twist shadowing related to the nonzero $\bar cc$ separation does
exist \cite{kt-charm}, but is much weaker \cite{kt-charm}. The leading
twist gluon shadowing is also rather weak even at the RHIC energies. As
for higher twist corrections, in the potential model \cite{kt-charm} they
may be large, but as we mentioned, this is not a solid theoretical
prediction.

Note that our description of nuclear suppression of heavy quarkonia is
quite different from the model proposed in \cite{bckt}. That model
involves three unknown parameters fitted to the nuclear data to be
explained. The key parameter is the absorption cross section for a dipole
consisting of a colored heavy quark pair $\bar QQ$ and light quarks.
However, the pair of heavy quarks that eventually forms the detected
quarkonium, and the comoving light quarks, cannot "talk to each other" due
to Lorentz time dilation during propagation through the nucleus. In other
words, it does not make any difference whether the accompanying light
quarks are primordial or are created during hadronization of the
color-octet $\bar QQ$ pair. Therefore, the multiple interactions of such a
large dipole should not be treated as absorption for production of a
colorless $\bar QQ$ pair via color neutralization. In contrast, in our
approach the strong suppression of heavy quarks is related to the steep
$z$-dependence of the fragmentation function $D_{\Psi/q}(z)$. This
function is fitted to data on $pp$ collisions, while no fitting is done to
nuclear data.

\section{Conclusions and outlook}

Nuclei suppress the large $x_F$ production of different species of
particles: light hadrons of both small and large $p_T$, dileptons, hidden
and open heavy flavor, photons, etc. So far, no exception is known. The
source of the effect can be understood either as an extra Sudakov
suppression caused by multiple interactions in nuclei, or as a
nucleus-induced reduction of the survival probability of large rapidity
gap processes. Although this effect can be also represented as an
effective energy loss, the former scales in $x_F$. This is different from
a simple single-parton energy loss, which is energy independent and leads
to an energy shift $\Delta x_1=\Delta E/E$ that vanishes with energy.

In addition to this key observation, the new results of this paper can be
presented as follows.

\begin{itemize}

\item The simple formula (\ref{120}) based on Glauber-Gribov multiple
interaction theory and the AGK cutting rules explains the universal $x_F$
scaling observed in data for inclusive production of leading light hadrons
with small $p_T$ quite well.

 \item With the same input we calculated high-$p_T$ hadron production at
large $x_F$ and found a substantial suppression. This parameter-free
calculation agrees with recent measurements performed by the BRAHMS
collaboration at forward rapidities in deuteron-gold collisions at RHIC.
Our simple explanation is based on just energy conservation; therefore, it
could be implemented independently of the dynamics. On the other hand, it
does not leave much room for other mechanisms under debate, such as the
CGC. We expect a similar suppression at large $p_T$ and large $x_F$ at
lower energies, where no effect of coherence is possible.

\item The Drell-Yan process, treated like heavy photon bremsstrahlung in
the target rest frame, is also subject to a nuclear suppression at large
$x_F$ imposed by energy conservation restrictions. In this case, we made a
model-dependent calculation and relied on a fit to data for soft
production of light hadrons. Within experimental uncertainties of the
available data, we described rather well the data for DY reaction at large
masses where an alternative explanation, nuclear shadowing, is excluded.

\item Charmonium production is different from DY only by a steeper
fragmentation function $q\to \Psi q$, which we fitted to $pp$ data, and by
an additional contribution of higher twist shadowing related to the large
size of the charmonium. We correctly reproduced the magnitude of nuclear
suppression at large $x_F$, but the shape of the $x_F$ dependence needs to
be improved. This problem seems to be a result of our model-independent,
but oversimplified, fit to soft hadronic data. We leave this improvement
for future work, both for charmonium production and the DY process.

\end{itemize}

\vspace*{0cm}

\begin{acknowledgments}

We are grateful to Stan Brodsky, Gerry Garvey, J\"org H\"ufner, Alexey
Kaidalov, Yuri Kovchegov, Mike Leitch, Genya Levin, Pat McGaughey,
Hans-J\"urgen Pirner, and J\"org Raufeisen for useful discussions and to
Berndt M\"uller for pointing to typos. The work of J.N. has been supported
by the Slovak Funding Agency, Grant No. 2/4063/24. Work was supported in
part by Fondecyt (Chile) grants 1030355, 1050519 and 1050589.

\end{acknowledgments}


\begin{thebibliography}{99}

\bibitem{brahms} BRAHMS Collaboration, I.~Arsene et al.,
Phys. Rev. Lett. {\bf 93}, 242303 (2004).

\bibitem{e-loss} M.B.~Johnson et al., Phys. Rev. Lett. {\bf 86}, 4483
(2001); Phys. Rev. C {\bf 65}, 025203 (2002).

\bibitem{maor} E.~Gotsman, E.M.~Levin and U.~Maor, Phys. Rev. D{\bf 60},
094011 (1999).

\bibitem{bb} R.~Blankenbecler and S.J.~Brodsky, Phys.Rev. D {\bf 10}, 2973
(1974).

\bibitem{bf} S.J.~Brodsky and G.R.~Farrar, Phys. Rev. Lett. {\bf 31}, 1153
(1973).

\bibitem{bs} I.A.~Schmidt and R.~Blankenbecler, Phys. Rev. D {\bf 15},
3321 (1977).

\bibitem{n} F. Niedermayer, Phys. Rev. D {\bf 34}, 3494 (1986).

\bibitem{bh} S.J.~Brodsky and P.~Hoyer, Phys. Lett. B {\bf 298}, 165 (1993).

\bibitem{bdms} R.~Baier, Yu.L.~Dokshitzer, A.H.~Mueller, S. Peigne and
D.~Schiff, Nucl. Phys. B {\bf 483}, 297 (1997); B {\bf 484}, 265 (1997).

\bibitem{knp} B.Z.~Kopeliovich, J.~Nemchik and E.~Predazzi, Proceedings
of the workshop on Future Physics at HERA, ed. by G.~Ingelman,
A.~De~Roeck and R.~Klanner, DESY 1995/1996, v. 2, 1038 (nucl-th/9607036);
Proceedings of the ELFE Summer School on Confinement physics, ed. by
S.D.~Bass and P.A.M.~Guichon, Cambridge 1995, Editions Frontieres, p. 391
(hep-ph/9511214).

\bibitem{knph} B.Z.~Kopeliovich, J.~Nemchik, E.~Predazzi, and
A.~Hayashigaki, Nucl. Phys. A {\bf 740}, 211 (2004).

\bibitem{hoyer}  P.~Hoyer, M.~Vanttinen, and U.~Sukhatme,
Phys. Lett. B {\bf 246}, 217 (1990).

\bibitem{factorization} J.~C.~Collins, D.E.~Soper, and G.~Sterman,
Nucl. Phys. B{\bf 261}, 104 (1985); hep-ph/0409313.

\bibitem{hir} B.Z. Kopeliovich {\sl Soft Component of Hard Reactions and
Nuclear Shadowing (DIS, Drell-Yan reaction, heavy quark production)}, in
proc. of the Workshop 'Dynamical Properties of Hadrons in Nuclear
Matter', Hirschegg 1995, ed. H. Feldmeier and W. Noerenberg, p. 102
(hep-ph/9609385).

\bibitem{glr} L.V.~Gribov, E.M.~Levin and M.G.~Ryskin, Nucl. Phys.
B {\bf 188}, 555 (1981); Phys. Rep. {\bf 100},1 (1983).

\bibitem{al} A.H.~Mueller, Eur. Phys. J. A {\bf 1}, 19 (1998).

\bibitem{mv} L.~McLerran and R.~Venugopalan, Phys. Rev. D {\bf 49}, 2233
(1994); D {\bf 49}, 3352 (1994).

\bibitem{kkt} D.~Kharzeev, Y.V.~Kovchegov and K.~Tuchin, Phys. Lett. B {\bf
599}, 23 (2004).

\bibitem{agk} A.V.~Abramovsky, V.N.~Gribov and O.V.~Kancheli, Yad. Fiz.
{\bf 18}, 595 (1973).

\bibitem{barton} D.S.~Barton et al., Phys. Rev. D {\bf 27}, 2580 (1983).

\bibitem{geist} W.M.~Geist, Nucl. Phys. A {\bf 525}, 149c (1991).

\bibitem{data}
D.S.~Barton et al., Phys. Rev. D {\bf 27}, 2580 (1983);
A.~Beretvas et al., Phys. Rev. D {\bf 34}, 53 (1986);
M.~Binkley et al., Phys. Rev. Lett. {\bf 37}, 571 (1976);
R.~Bailey et al., Z. Phys. C {\bf 22}, 125 (1984);
P.~Skubic et al., Phys. Rev. D {\bf 18}, 3115 (1978).

\bibitem{e866-psi} The E866 Collaboration, M.J.~Leitch et al., Phys. Rev.
Lett. {\bf 84}, 3256 (2000).

\bibitem{na3} The NA3 Collaboration, J.~Badier et al., Z. Phys. C {\bf 20},
101 (1983).

\bibitem{psi-qm} PHENIX Collaboration, R.G.~de~Cassagnac, talk at Quark
Matter, Berkeley, 2004.

\bibitem{gb} J.F.~Gunion and G.~Bertsch, Phys. Rev. D {\bf 25}, 746 (1982).

\bibitem{gribov} V.N.~Gribov, Eur. Phys. J. C {\bf 10}, 71 (1999).

\bibitem{k3p} B.Z.~Kopeliovich, I.K.~Potashnikova, B.~Povh, and E.~Predazzi,
Phys. Rev. Lett. {\bf 85}, 507 (2000); Phys. Rev. D {\bf 63}, 054001 (2001).

\bibitem{kaidalov} A.B.~Kaidalov, JETP Lett. {\bf 32}, 474 (1980); Sov. J.
Nucl. Phys. {\bf 33}, 733 (1981); Phys. Lett. B {\bf 116}, 459 (1982).

\bibitem{capella} A.~Capella et al., Phys. Rep. {\bf 236}, 225 (1994).

\bibitem{mine} B.Z.~Kopeliovich, Phys. Rev. C {\bf 68}, 044906 (2003).

\bibitem{helios} T. Akesson et al., Z. Phys. C {\bf 49}, 355 (1991).

\bibitem{kps} B.Z.~Kopeliovich, I.K.~Potashnikova and I.A.~Schmidt,
{\sl Large rapidity gap events in proton-nucleus collisions}, paper in
preparation.

\bibitem{bckt} K.~Boreskov, A.~Capella, A.~Kaidalov and J.~Tran~Thanh~Van,
Phys. Rev. D {\bf 47}, 919 (1993).

\bibitem{cronin} D.~Antreasyan et al., Phys. Rev. D {\bf 19}, 764 (1979).

\bibitem{knst} B.Z.~Kopeliovich, J.~Nemchik, A.~Sch\"afer and
A.V.~Tarasov, Phys. Rev. Lett. {\bf 88}, 232303 (2002).

\bibitem{mytalk} B.Z.~Kopeliovich, talk at the Workshp on High-$p_T$
Physics at RHIC, December 2-6, 2003.

\bibitem{grv} M.~Gluck, E.~Reya and A.~Vogt, Z. Phys. C {\bf 67}, 433
(1995).

\bibitem{zkl} A.B.~Zamolodchikov, B.Z.~Kopeliovich and L.I.~Lapidus, Sov.
Phys. JETP Lett. {\bf 33}, 595 (1981); Pisma v Zh. Exper. Teor.  Fiz. {\bf
33}, 612 (1981).

\bibitem{jkt} M.B.~Johnson, B.Z.~Kopeliovich and A.V.~Tarasov, Phys. Rev.
C {\bf 63}, 035203 (2001).

\bibitem{review} M.~Anselmino et al., Rev. Mod. Phys. {\bf 65}, 1199 (1993).

\bibitem{kst1} B.Z.~Kopeliovich, A.~Sch\"afer and A.V.~Tarasov,
Phys. Rev. C {\bf 59}, 1609 (1999).

\bibitem{kst2} B.Z.~Kopeliovich, A.~Sch\"afer and A.V.~Tarasov, Phys.
Rev. D {\bf 62}, 054022 (2000).

\bibitem{gbw} K.~Golec-Biernat and M.~W\"usthoff, Phys. Rev. D {\bf 59}
014017 (1999).

\bibitem{3f} R.P.~Feynman, R.D.~Field, and G.C.~Fox, Phys. Rev. D {\bf 18},
3320 (1978).

\bibitem{km} Yu.V.~Kovchegov and A.H.~Mueller, Nucl.  Phys. B {\bf 529}, 451
(1998).

\bibitem{florian} D.~de~Florian and R.~Sassot, Phys. Rev. D {\bf 69},
074028 (2004).

\bibitem{hermes} S.~Kretzer, E.~Leader and E.~Christova, Eur. Phys. J. C
{\bf 22}, 269 (2001).

\bibitem{kp}  B.Z.~Kopeliovich, B.~Povh, J. Phys. G {\bf 30}, S999 (2004).

\bibitem{pisa} M.~D'Elia, A.~Di~Giacomo and E.~Meggiolaro,
Phys. Lett. B {\bf 408}, 315 (1997).

\bibitem{shuryak} E.V.~Shuryak, I.~Zahed, Phys. Rev. D {\bf 69}, 014011
(2004).

\bibitem{e772} D.M.~Alde et al., Phys. Rev. Lett. {\bf 64}, 2479 (1990).

\bibitem{kn1983} B.Z. Kopeliovich and F. Niedermayer, {\sl
Nuclear screening in $J/\Psi$ and Drell-Yan pair production},
JINR-E2-84-834, Dubna 1984 (scanned in KEK library).

\bibitem{vasiliev} M. Vasiliev et al., Phys. Rev. Lett. {\bf 83}, 2304
(1999).

\bibitem{bhq}  S.J.~Brodsky, A.~Hebecker and  E.~Quack,
 Phys. Rev. D {\bf 55}, 2584 (1997).

\bibitem{krt3} B.Z.~Kopeliovich, J.~Raufeisen and A.V.~Tarasov, Phys. Lett.
B {\bf 503}, 91 (2001).

\bibitem{krtj} B.Z.~Kopeliovich, J.~Raufeisen, A.V.~Tarasov, and
M.B.~Johnson, Phys. Rev. C {\bf 67}, 014903 (2003).

\bibitem{bms} R.~Baier, A.H.~Mueller, D.~Schiff, Nucl. Phys. A {\bf 741}
358 (2004).

\bibitem{steq}  J. Pumplin et al., JHEP {\bf 0207}, 012 (2002).

\bibitem{katsanevas} E537 Collaboration, S.~Katsanevas et al., Phys. Rev.
Lett. {\bf 60}, 2121 (1988).

\bibitem{kth} B.Z.~Kopeliovich, A.V.~Tarasov, and J.~H\"ufner, Nucl. Phys.
A {\bf 696}, 669 (2001).

\bibitem{kz1991} B.Z.~Kopeliovich and B.G.~Zakharov, Phys. Rev. D {\bf 44},
3466 (1991).

\bibitem{hk-lc} J.~H\"ufner, and B.Z.~Kopeliovich, Phys. Lett. B {\bf 403},
128 (1997).

\bibitem{hikt2} J.~H\"ufner, Yu.P.~Ivanov, B.Z.~Kopeliovich and
A.V.~Tarasov, Phys. Rev. C {\bf 66}, 024903 (2002).

\bibitem{hkp} J.~H\"ufner, B.Z.~Kopeliovich and A.~Polleri, Phys. Rev. Lett.
{\bf 87}, 112302 (2001).

\bibitem{krt2} B.Z.~Kopeliovich, J.~Raufeisen and A.V.~Tarasov, Phys. Rev.
C {\bf 62}, 035204 (2000).

\bibitem{kt-charm} B.Z.~Kopeliovich, A.V.~Tarasov, Nucl. Phys. A {\bf
710}, 180 (2002).

\end{thebibliography}
\end{document}